\documentclass[]{article}
\usepackage[a4paper,left=2.5cm,right=2.5cm,top=3cm,bottom=3cm]{geometry}
\usepackage{graphicx}
\usepackage{bm}
\usepackage{amsmath}
\usepackage{amsfonts} 
\usepackage{amssymb}
\usepackage[colorlinks=true,allcolors=blue]{hyperref}
\usepackage{float}
\usepackage[section]{placeins}
\usepackage[numbers,sort&compress]{natbib} 
\begin{document}

\title{Evaluating Local Viscoelastic Properties of Artery-mimicking Phantoms using High-Resolution Flexural Wavenumber Analysis}%

\date{October 2026}
\author{Max Gattin, Julien de Rosny\\[4pt]
\small ESPCI Paris, PSL University, Institut Langevin, CNRS, 75005 Paris, France}

\maketitle

\begin{abstract}
The measurement of guided wave propagation is an effective approach for evaluating arterial wall stiffness. A more comprehensive characterization of viscoelastic material properties could yield more informative biomarkers; however, this additionally requires estimation of wave attenuation. Furthermore, spatially local estimation is necessary for localized pathologies, such as atherosclerotic plaques. To this end, we propose an original two-stage high-resolution method to recover local viscoelastic properties from guided waves in artery-like soft tubes. First, the local $f$--$k$ dispersion relation including losses is estimated using a high-resolution complex wavenumber identification method. Second, assuming a fractional Kelvin–Voigt rheological model for the tube wall, we estimate the viscoelastic material parameters by fitting the dispersion relation computed with a spectral element method to this estimate. We experimentally validate the proposed approach on flexural wave propagation along water-filled soft tubes, measured with a high-speed camera coupled with digital image correlation. Owing to the high resolution of the method, the axial modulus variation of a graded phantom is recovered. Finally, a Cramér–Rao analysis indicates that the complex wavenumber estimator operates near its theoretical limit, and reveals why local attenuation is intrinsically difficult to recover. 
\end{abstract}

\section*{\label{sec:intro} Introduction}

Arterial wall mechanical properties are established markers of cardiovascular risk, reflecting the structural changes associated with aging, hypertension, and atherosclerosis~\cite{laurentExpertConsensusDocument2006,vlachopoulosPredictionCardiovascularEvents2010}. Carotid--femoral pulse wave velocity (cfPWV) is the clinical reference for arterial stiffness assessment: it is well validated~\cite{ben-shlomoAorticPulseWave2014}, but it provides only a \emph{global} estimate along a long arterial segment and cannot resolve variations associated with \emph{local} disease such as atheroma plaques. This has motivated a sustained effort toward \emph{local} evaluation of arterial elasticity.

Most local approaches ultimately aim to recover the wall stiffness, and differ in how they do so~\cite{dekorteReviewMechanicalCharacterization2016}. A first family of methods estimates local stiffness based on quasi-static mechanical deformation assumptions, either through a stress-strain analysis that uses the natural pulse wave to generate local stress~\cite{schmittNoninvasiveVascularElastography2007,naimCharacterisationCarotidPlaques2013}, or based on the amplitude of tissue displacement measured at the focus of an acoustic radiation force (ARFI)~\cite {dahlAcousticRadiationForce2009, czernuszewiczPerformanceAcousticRadiation2017}. A second family instead exploits the propagation of mechanical waves along the wall, whose velocity is related to the stiffness. Depending on the wave generation, different mechanical waves are observed. In the \emph{active} approach, an acoustic radiation force launches shear waves that are then tracked, up to a frequency of 2~kHz; these methods have been validated from tissue-mimicking phantoms~\cite{couadeQuantitativeAssessmentArterial2010, bernalMaterialPropertyEstimation2011, maksutiArterialStiffnessEstimation2016, widmanShearWaveElastography2016, royMultimodalGuidedWave2021} through \emph{ex vivo} arteries~\cite{widmanShearWaveElastography2016} to \emph{in vivo} carotids~\cite{couadeQuantitativeAssessmentArterial2010,jiangSimultaneousImagingBidirectional2025}. In the \emph{passive} approach, the natural pulse wave driven by the heartbeat is tracked at the low-frequency spectral content of the cardiac cycle, giving the local pulse-wave velocity \emph{in vivo}~\cite{apostolakisPiecewisePulseWave2016,maraisArterialStiffnessAssessment2019}. While no external source is required, the longer wavelengths involved, combined with the limited field of view of ultrasound imaging systems, make this passive approach more challenging. Beyond ultrasound, this passive approach has been explored with other modalities, including laser Doppler holography in retinal arteries~\cite{laloy-borgnaObservationNaturalFlexural2023}, x-rays in coronary arteries~\cite{gregoireQuantitativeXrayElastography2025}, and optical coherence elastography in excised porcine aorta~\cite{jiangComprehensiveCharacterizationNonlinear2026}. 

Two recent developments motivate the present work. The first is the prospect of \emph{viscoelastic} characterization to recover not only the wall stiffness but also its frequency-dependent, dissipative behavior as a complementary biomarker. Until very recently~\cite{roy2026frequency}, this prospect was confined to simulation~\cite{royFullWaveformInversion2023} and to high-frequency capabilities of optical coherence elastography~\cite{jiangComprehensiveCharacterizationNonlinear2026}. The second is the emergence of \emph{flexural} waves, recently shown to be carried by the natural low-frequency wavefield in the arterial wall~\cite{laloy-borgnaObservationNaturalFlexural2023,gregoireFlexuralPulseWave2024,gregoireQuantitativeXrayElastography2025}. Due to their dispersive nature and relatively low phase velocity, they seem to possess characteristics that are particularly well-suited for quantitative measurement, yet they remain comparatively underexplored. 

These propagation-based approaches all rely on estimating the dispersion relation of the resulting waveguide, which serves as a fingerprint of the material’s properties. The material parameters are then obtained by matching this observed dispersion relation to that predicted by a waveguide model. However, the determination of viscoelastic properties is intrinsically linked to the \textit{imaginary part} of the dispersion relation, \textit{i.e.} to attenuation, which is more difficult to determine experimentally than the real part alone. Furthermore, the local resolution of these properties introduces an additional, independent constraint: whether due to a limited field of view or the inherently localized nature of the clinical features of interest, this requires working with a limited spatial segments, typically smaller than or comparable to the wavelength.

Indeed, the standard tool for revealing the dispersion relation, across fields, is the two-dimensional Fourier transform, which maps a spatio-temporal measured field into its frequency–wavenumber representation. It fails on both points just identified: it returns only the real part of the wavenumber, omitting the attenuation entirely, and by the Rayleigh criterion it requires a spatial domain that is large compared to the wavelength, which is in direct contradiction to the local estimate required here. Attenuation estimation is also typically only achieved under restrictive assumptions, such as a single forward-propagating plane wave or a known resonant structure. A different strategy consists of entirely bypassing the dispersion relation as an intermediate quantity by directly fitting the results of a numerically solved propagation model to the measured field in order to determine the viscoelastic parameters. This technique is sometimes referred to as full waveform inversion (FWI) \cite{royFullWaveformInversion2023}. However, the reliability of such an approach depends heavily on the validity of the chosen model. This is precisely the problem when it comes to blood vessels, for which modeling wave propagation is difficult and often requires important simplifying assumptions. Another strategy for measuring attenuation is based on signal subspace methods, including Matrix-Pencil, MUSIC (MUltiple SIgnal Classification) and ESPRIT algorithm (Estimation of Signal Parameters via Rotational Invariance Techniques) adopted here~\cite{margeritExtractionDispersionSurfaces2022}. 
This method aims to identify a sparse sum of complex exponentials in background noise, which offers four notable advantages over the approaches mentioned above: (i) it allows for the recovery of both the real and imaginary parts of the wavenumber, (ii) it allows for the disentanglement of multiple modes and backward propagation, (iii) it operates beyond the resolution limit of Fourier-transform-based approaches, and most importantly (iv) it requires no prior knowledge of the constitutive equation of the waveguide or the rheological model. Of these methods, and to the authors’ knowledge, only the Matrix-Pencil technique has recently been used in the context of ultrasonic viscoelastography for noise reduction~\cite{liDelayedMatrixPencil2024}.

Here, we propose a two-stage method to first recover the high-resolution complex dispersion relation as a model-agnostic observable, then apply a model-based inverse problem to recover viscoelastic properties. The approach is demonstrated experimentally with flexural waves measured on custom-made soft tubular phantoms filled with water. Section~\ref{sec:model} introduces the guided-wave model of a fluid-filled viscoelastic tube and the computation of its complex dispersion relation using a preexisting spectral element method. Section~\ref{sec:method} presents the two-stage method for viscoelastic characterization, first introducing the mathematical foundations of ESPRIT for estimating complex dispersion relations in one-dimensional waveguides, then proposing a model-based inverse problem formulation to recover the wall's viscoelastic properties from the complex dispersion curve thus obtained. Section~\ref{sec:experimental methods} describes the fabrication of custom-made silicone tube phantoms, including a sample with spatially graded properties to demonstrate the local/super-resolution capability of the proposed method, as well as the setup for generating and acquiring the wavefield using a high-speed camera and digital image correlation. Section~\ref{sec:results} presents the experimental results and shows how the method can be used to obtain a local estimate of the viscoelastic properties. Finally, Section~\ref{sec:uncertainty} quantifies the accuracy that can be achieved with this type of method and analyzes some of the fundamental limitations that make local viscoelastic measurements structurally challenging.

\section{\label{sec:model} Guided wave model of a fluid-filled viscoelastic tube}

A fluid-filled soft tube of inner radius $r$ and wall thickness $h$ is proposed as a simplified model of a blood vessel. The inner fluid is characterized by its mass density $\rho_f$ and pressure wave velocity $c_f$. The wall material is modeled as a quasi-incompressible elastic solid with Poisson's ratio $\nu = 0.499995$, consistent with the nearly incompressible behavior of soft biological tissues. The wall mass density is $\rho$, and its viscoelastic behavior is described by a complex, frequency-dependent Young modulus $E(\omega)$. We decided to parametrize it with a fractional Kelvin--Voigt model, a widely used phenomenological model for soft biological tissues and soft polymers~\cite{delorySoftElastomersPlayground2022,croquetteGuidedElasticWaves2026,jiangComprehensiveCharacterizationNonlinear2026,liuUltraWideband05162014}:
\begin{equation} \label{eq:voigt}
    E(\omega) = E_0 \left(1 + (i\omega\tau)^n\right)\quad,
\end{equation}
where $E_0$ is the static Young modulus, $\tau$ is the relaxation time, and $n \in (0,1]$ is the fractional exponent. Here $i=\sqrt{-1}$ is the imaginary unit and $\omega = 2\pi f$ is the angular frequency.

Dispersion relations are computed using the open-source GEWtool MATLAB code~\cite{kieferGEWtoolComputerSoftware2023}, which solves the boundary value problem over the waveguide cross-section using a high-order spectral element method, thereby enabling a relatively short computation time. Unlike some other dispersion relation solvers that require a \emph{real} wavenumber $k$ to calculate the associated \emph{real} eigenfrequencies $\omega$, GEWtool allows to specify a \emph{real} frequency $\omega$ to calculate the associated \emph{complex} wavenumber, whose imaginary part directly encodes the attenuation of the associated mode. Thanks to the mode decoupling in axisymmetric cylindrical waveguides, GEWtool solves the dispersion relations separately for three kinds of uncoupled solutions: longitudinal $L(0,q)$, torsional $T(0,q)$, and flexural $F(m,q)$, where $m$ is the circumferential order and $q$ is the branch order.

\begin{figure}[htbp]
    \centering
    \includegraphics[width=0.6\linewidth]{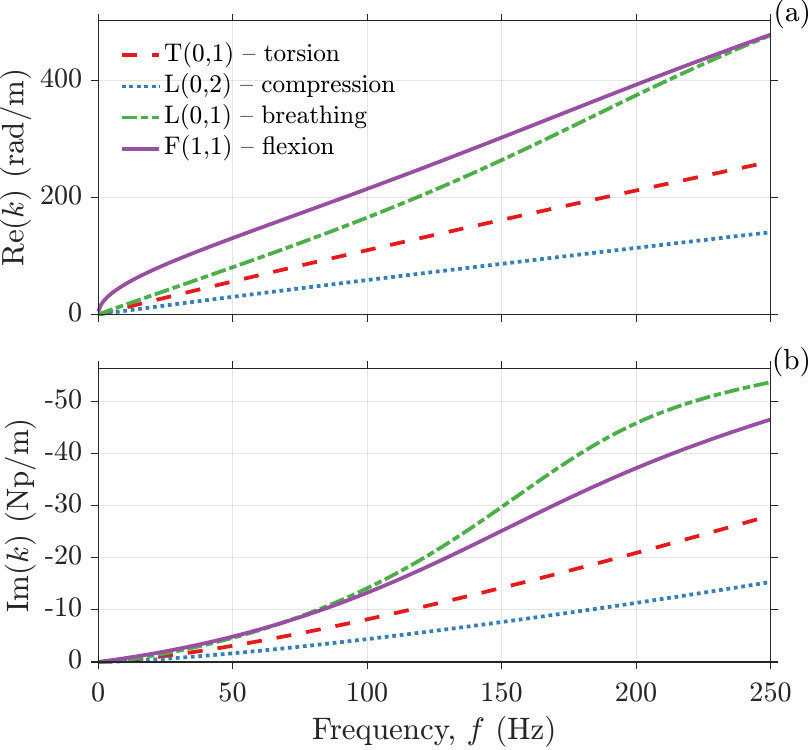}
    \caption{(a) Real and (b) imaginary parts of the dispersion relation of a fluid-filled soft tube ($r=4$~mm, $h=2$~mm, $\rho_f=1000$~kg/m$^3$, $c_f=1500$~m/s, $c_s=10^{-3}$~m/s, $\rho=1080$~kg/m$^3$, $E_0=90$~kPa, $\tau=100$~$\mu$s, $n=0.5$).}
    \label{fig:model_dispersion}
\end{figure}

 Figure~\ref{fig:model_dispersion} shows an example of the complex dispersion relation for the first few fundamental branches (\textit{i.e.} modes for which the cutoff frequency is zero), computed for a representative parameter set (see caption). Note that GEWtool computes both positive and negative branches (forward and backward waves). As they are identical apart from the sign, only the branch with a positive real part is shown here. Furthermore, a small but finite shear wave velocity $c_s$ is assigned to the fluid domain, as GEWtool does not support explicit fluid modeling; its specific value has a negligible impact on the computed dispersion relation.

\begin{figure*}[htbp]
    \centering
    \includegraphics[width=0.8\linewidth]{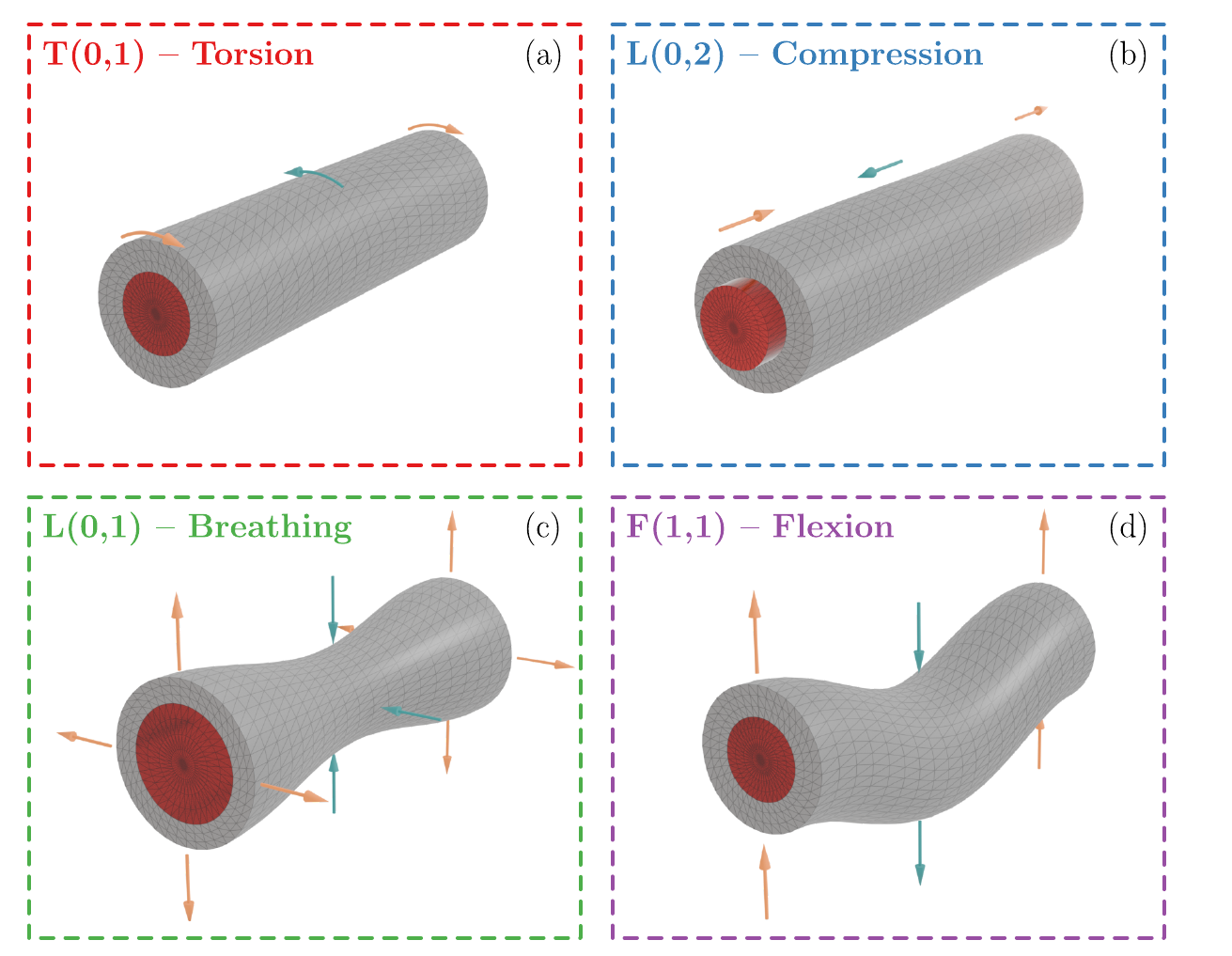}
    \caption{Schematic representation of the four fundamental modes of fluid-filled tubes: (a) torsional $T(0,1)$, (b) wall-dominated longitudinal $L(0,2)$, (c) fluid-dominated longitudinal (\textit{breathing}) $L(0,1)$, and (d) flexural $F(1,1)$ modes.}
    \label{fig:tube_modes}
\end{figure*}

In the low-frequency range under consideration, four guided modes are present whose complex dispersion relations are represented in Figure~\ref{fig:model_dispersion}. The displacement fields associated with each one of them are schematically represented in Figure~\ref{fig:tube_modes}: the torsional mode $T(0,1)$ (a) is axisymmetric and polarized purely azimuthally. It is present almost only in the wall, as the fluid does not carry shear-polarized waves. For a fluid-filled tube, two longitudinal mods coexist, both being axisymmetric, polarized axially and radially: a wall-dominated branch $L(0,2)$ (b), close to a free rod's compressional wave, and a fluid-dominated branch $L(0,1)$ (c), corresponding to the radial expansion and contraction of the wall driven by the internal pressure, classically termed ``breathing''  mode. Finally, the flexural mode $F(1,1)$ (d) is the only non-axisymmetric mode ($m\neq 0)$, polarized axially, radially, and azimuthally; it corresponds to the bending motion of the tube. It is also markedly different from the other three: it is much slower and highly dispersive in the very low-frequency regime, its phase velocity varying with frequency and scaling as $\sqrt{\omega}$ (\textit{i.e.} phase and group velocity tend to 0 in the quasi-static regime)~\cite{gregoireFlexuralPulseWave2024}. 

To the best of the authors' knowledge, neither wall-dominated longitudinal nor torsional waves have been measured \emph{in vivo} in blood vessels, and it is even unclear how they would be excited or measured in practice, as they carry little to no radial displacement. The breathing mode is the predominant mode of the arterial pulse wave naturally excited by the heartbeat; its low-frequency velocity is approximated by the Moens--Korteweg relation, and it is the mode classically exploited for pulse-wave-velocity measurements~\cite{dekorteReviewMechanicalCharacterization2016}. Flexural waves have only recently been shown to be observable \emph{in vivo}~\cite{laloy-borgnaObservationNaturalFlexural2023,gregoireQuantitativeXrayElastography2025}. Being dispersive, slow, and with a radial displacement component, they are the mode most appropriate for measuring wave attenuation, and hence for recovering the wall's viscoelastic properties. For these reasons, the remainder of this article will focus specifically on this fundamental flexural mode. Note that only the fundamental branches of each mode family are considered here. There are indeed also higher-order solutions ($m{>}1$ or $q{>}1$), which are correctly predicted by the model but are deliberately not shown here, as they are evanescent over the frequency range of interest.

\section{\label{sec:method} A two-stage method for viscoelasticity estimation}

\begin{figure*}[htbp]
    \centering
    \includegraphics[width=0.99\linewidth]{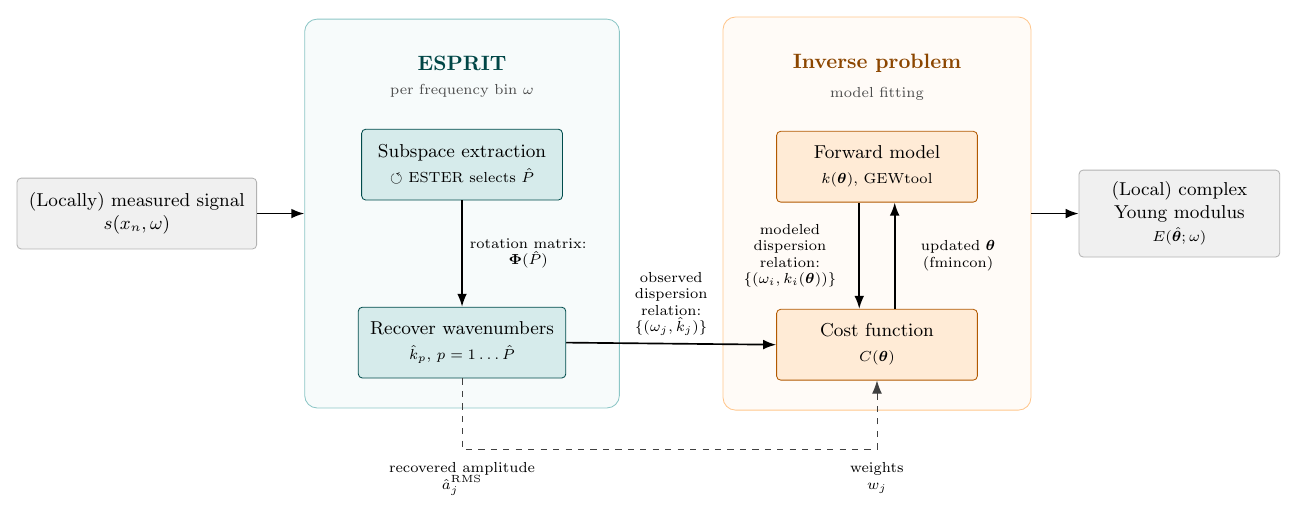}
    \caption{Flowchart of the two-stage viscoelastic parameter estimation method.}
    \label{fig:flowcharts}
\end{figure*}

A given  dispersion relationship can be regarded as an encoded measure of the mechanical properties, assuming that the other parameters are known (here, wall thickness $h$, tube inner radius $r$, wall density $\rho$, fluid density $\rho_f$ and fluid phase velocity $c_f$), at least to some degree of uncertainty. Consequently, a commonly used strategy for determining viscoelastic properties from the dispersion relation involves solving an inverse problem to find the model parameters that provide the best correspondence with the experimentally observed dispersion curve. The approach proposed here breaks down the inverse problem into two stages, as illustrated in Figure~\ref{fig:flowcharts}. First, a high-resolution wavenumber analysis method, based on the ESPRIT algorithm, is applied to estimate the complex dispersion relation from the measured spatiotemporal signal. Only then, a model-based inverse problem is solved to determine the rheological parameters by matching the modeled complex dispersion relation to the one estimated with ESPRIT.

\subsection{\label{subsec:method:esprit} High-resolution extraction of the complex dispersion curves using ESPRIT}

The ESPRIT algorithm and similar methods such as Matrix-Pencil belong to the family of high-resolution subspace techniques aimed at estimating a sum of discrete damped sinusoids in the presence of background noise. Originally developed for the Direction-Of-Arrival (DOA) estimation problem for array signal processing \cite{royESPRITestimationSignalParameters1989}, it has since been applied to elastic waveguide problems to recover complex wavenumbers~\cite{margeritHighResolutionWavenumberAnalysis2018,margeritHighResolutionWavevectorAnalysis2019, changExtractionGuidedWave2018,despres2024characterization}.

ESPRIT relies on two physical assumptions. The \textit{sparsity} assumption requires that the number of propagating modes $P$ is small compared to the number of measurement points $N$. The \textit{rotational invariance} assumption requires that the wavefield decomposes as a superposition of plane waves with complex wavenumbers, measured at regularly spaced positions. Both are easily satisfied and well justified in the context of 1D guided wave propagation. By explicitly incorporating these prior assumptions, ESPRIT achieves spectral resolution beyond the classical Fourier limit (Rayleigh criterion) without requiring a longer measurement aperture.

Consider the complex monochromatic wavefield measured at $N$ equally spaced positions $x_n = n\Delta x$, $n=1,\ldots,N$, at angular frequency $\omega$. Assuming $P$ propagating modes, the signal at position $x_n$ is modeled as:
\begin{equation} \label{eq:signal_model}
s(x_n, \omega) = \sum_{p=1}^{P} a_p(\omega) e^{ik_p(\omega) x_n} + b(x_n, \omega) \quad,
\end{equation}
where $a_p(\omega)$ is the complex amplitude of the  $p$-th mode, $k_p(\omega) = k_p'(\omega) +i k_p''(\omega) \in \mathbb{C}$ is its complex wavenumber whose imaginary part encodes spatial attenuation, and $b(x_n,\omega) \in \mathbb{C}$ is a complex additive and spatially uncorrelated noise. For the sake of simplicity, frequency dependence is omitted hereafter. In matrix form:
\begin{equation}\label{signal_model_matrix}
\bm{s} = \bm{V}\bm{a} + \bm{b} \quad ,
\end{equation}
where $\bm{s} \in \mathbb{C}^N$ is the measured signal vector and $\bm{V} \in \mathbb{C}^{N\times P}$ is a Vandermonde matrix whose $p$-th column is:
\begin{equation}\label{Vandermonde}
\bm{V}_p = [z_p, z_p^2, \ldots, z_p^{N}]^T, \quad z_p = e^{ik_p\Delta x} \quad .
\end{equation}
The Vandermonde structure reflects the fact that each mode accumulates a constant complex phase shift and attenuation through $z_p$ coefficient between successive measurement points. This is the \textit{rotational invariance} property, which is the geometric foundation of ESPRIT. It can be expressed using sub-matrices obtained by removing the last row ($\bm{V}^-$) and the first row ($\bm{V}^+$) of $\bm{V}$ that satisfy:
\begin{equation}
\label{eq:rotationnal_invariance}
\bm{V}^+ = \bm{V}^- \bm{Z}, \quad \bm{Z} = \text{diag}(z_1,\ldots,z_P) \quad .
\end{equation}

Now consider the signal matrix $\bm{S}\in \mathbb{C}^{N\times M}$  formed with $M$ independent measurements, $\bm{S} = [\bm{s}^{(1)}, \ldots, \bm{s}^{(M)}]$. Thanks to the \textit{sparsity} assumption, it can be decomposed via Singular Value Decomposition (SVD) into signal and noise subspaces. Retaining the $P$ dominant singular vectors yields a signal subspace matrix $\bm{U} \in \mathbb{C}^{N\times P}$ that spans approximately the same subspace as $\bm{V} \in \mathbb{C}^{N\times P}$, up to an invertible matrix $\bm{T} \in \mathbb{C}^{P\times P}$ which behaves as a change of basis operator:
\begin{equation}
\bm{V} = \bm{U}\bm{T}\quad.
\end{equation}
Substituting this last relation in the rotational invariance (Equation~\eqref{eq:rotationnal_invariance}), we can define the rotation matrix $\bm{\Phi}$ which naturally takes the form of an eigendecomposition problem:
\begin{equation}
\bm{\Phi} = [\bm{U}^-]^\dagger \bm{U}^+ = \bm{T}\bm{Z}\bm{T}^{-1} \quad ,
\end{equation}
where $^\dagger$ denotes the Moore--Penrose pseudo-inverse. The estimated $\hat{z}_p$ values are the eigenvalues of $\bm{\Phi}$, from which the complex wavenumbers are finally recovered as:
\begin{equation} \label{eq:k}
\hat{k}_p = \frac{1}{i\Delta x}\log(\hat{z}_p), \quad p = 1,\ldots,P \quad .
\end{equation}
The imaginary part of $\hat{k}_p$ gives the estimation of the spatial attenuation of the $p$-th mode directly, without any assumption on the waveguide besides that the measured field can be written as a sum of plane waves and uncorrelated additive noise.

In practice, forming a rank-$P$ signal matrix $\bm{S}$ for signal subspace identification is difficult as it would require performing multiple measurements with linearly independent amplitude vector $\bm{a}$, for example by using at least $P$ different sources. Indeed stacking linearly dependent measurements (for example multiple realizations with the same source) would only lead to a rank-$1$ signal matrix. For regularly spaced measurements, replacing the signal matrix $\bm{S}$ with a Hankel matrix $\bm{H}$ provides an elegant workaround. A single measurement vector $\bm{s}$ can be restructured into $K$ overlapping virtual realizations of length $L$, each shifted by one spatial step $\Delta x$:

\begin{equation}
\label{hankel}
    \bm{H} = \begin{bmatrix}
        s_1 & s_2 &... & s_{K} \\
        s_2 & s_3 &... & s_{K+1} \\
        ... & ... &\ddots & ... \\
        s_{L} & s_{L+1} &... & s_{N} \\
    \end{bmatrix} ~~  \in \mathbb{C}^{L \times K} \quad,
\end{equation}

with $K = N-L+1$. The choice of $L$ involves a trade-off: large $L$ preserves long-range spatial correlations and improves spectral resolution but increases sensitivity to noise, while small $L$ increases redundancy and improves noise robustness at the cost of discarding long range correlations and therefore degraded spectral resolution. In practice, it also impacts the performance of the subspace separation and $L \approx N/2$ appear to be a reasonable guideline~\cite{margeritHighResolutionWavenumberAnalysis2018}. Note also that the maximum number of recoverable modes is fundamentally bounded by $P \leq \min(K,L)$, as the maximum rank of the Hankel matrix cannot exceed its smallest dimension. The Hankel matrix formalism can be combined with multiple realization by computing the individual Hankel matrix of each realization $\bm{H}^{(m)}$ and simply stacking them together horizontally $\bm{H} = [\bm{H}^{(1)},\ldots, \bm{H}^{(M)}]$.

In general, the number of modes $P$ is also an unknown of the problem and must be estimated from the data as well. Here, we use the ESTER (ESTimation of ERror) criterion~\cite{badeauNewPerturbationAnalysis2006}, which evaluates the quality of the rotational invariance relation for each candidate integer value of $P$:
\begin{equation}
E(P) = \|\bm{U}^+(P) - \bm{\Phi}(P)\bm{U}^-(P)\|_2 \quad .
\end{equation}
The best estimate $\hat{P}$ is obtained by testing multiple values of $P$ and keeping the one that minimizes $E(P)$. A threshold variant of this criterion allows recovery of weaker modal components by retaining the largest values of $P$ for which $J(P) = 1/E(P)$ exceeds a fraction $\gamma$ of its maximum value, \textit{i.e.} $J(P)/\max(J(P)) \geq \gamma$. The choice of a specific $\gamma$ value then results from a trade-off between sensitivity to weak modes ($\gamma$ close to 0) and robustness to noise overfitting ($\gamma$ close to 1). 

As a final verification step, it is also possible to estimate a denoised version of the input signal. To do so, we first need to reconstruct an estimate of the Vandermonde matrix $\hat{\bm{V}}$ from equation \eqref{Vandermonde} given estimated wavenumbers $\hat{k}$. Then we can compute estimates of the complex amplitudes $\hat{\bm{a}}$ (in the least-squares sense) using Equation~\eqref{signal_model_matrix}:
\begin{equation}
\label{amplitude_vector_estimation}
    \hat{\bm{a}} = \hat{\bm{V}}^{\dagger}\bm{s} \quad .
\end{equation}
Incidentally, the denoised signal can also be reconstructed as:
\begin{equation}
\label{denoised_signal}
    \hat{\bm{s}} = \hat{\bm{V}}\hat{\bm{a}} \quad .
\end{equation}
Finally, for a more accurate representation of the total energy associated with each solution, we compute the Root-Mean-Square (RMS) amplitude over the spatial window of the denoised signal associated with each mode:
\begin{equation} \label{rms}
    \hat{a}_p^\text{rms} = \frac{1}{\sqrt{N}} \left\| \hat{a}_p \hat{\bm{v}}_p \right\|_2  \quad ,
\end{equation}
where $\hat{\bm{{v}}}_p$ and $\hat{a}_p$ are respectively the $p$-th column of the estimated Vandermonde matrix and the $p$-th value of the estimated complex amplitude vector.

ESPRIT offers four key benefits. First, unlike Fourier-based methods, ESPRIT the \emph{complex} wavenumber (Equation~\ref{eq:k}), which is essential information for determining viscoelastic properties. Second, thanks to \emph{sparsity} and \emph{rotationnal invariance} assumptions, ESPRIT operates beyond Rayleigh resolution limit, meaning that the dispersion relation can be estimated from limited spatial windows. Third, it allows to disentangle multiple modes and waves traveling in both directions, each being associated with one of the $P$ distinct eigenvalues. Finally, it is agnostic to the waveguide model and rheological model as it simply assumes that the field can be expressed as a sparse sum of damped plane waves (Equation~\ref{eq:signal_model}). Therefore, the estimated dispersion relation can be used as a standalone, \emph{model-independent} observable against which the validity of a given model can be critically evaluated. 

\subsection{\label{subsec:method:inverse} Model-based recovery of the complex modulus}

While the wavenumber-extraction stage above is agnostic to the underlying waveguide and material rheology, the recovery of the viscoelastic parameters necessarily commits to a specific choice for both. Here, we adopt the fluid-filled tube waveguide model and fractional Kelvin–Voigt rheology introduced in Sec.~\ref{sec:model}, parametrized with static Young modulus $E_0$, relaxation time $\tau$ and power-law exponent $n$. We consider only the modes from the flexural family, as longitudinal and torsional modes are not excited in our experimental configuration (Section~\ref{subsec:experimental methods:optical}). Fitting this model to the measured dispersion points raises two distinct difficulties. First, fitting a power-law model over a limited frequency range is known to be a poorly conditioned problem: a continuous range of parameter combinations can produce an almost equally convincing fit. In practice, for instance, a change in $n$ can be largely compensated by a change in $\tau$ and $E_0$. Second, the dispersion relation is generally multi-valued: a single frequency may correspond to several wavenumbers, each lying on a distinct branch. This happens when several propagating modes coexist, when the same mode propagates in both directions (forward and backward solutions), or when the model also predicts evanescent solutions. In practice, however, only some of these solutions are sufficiently excited to be detected, which complicates the process of correctly matching each observed dispersion relation branch with the predicted one.

To address the first difficulty, we rewrite the fractional Kelvin--Voigt model in the following frequency-normalized form (see Appendix~\ref{app:normalization}):
\begin{equation}
    E(\omega) = \Big[E_c' - E_c''\cot\big(n\pi/2\big)\Big]
    + \frac{E_c''}{\sin\big(n\pi/2\big)}
    \left(i\frac{\omega}{\omega_c}\right)^{n}
    \label{eq:voigt_normalized}
\end{equation}
where $E(\omega_c) = E_c = E_c' + iE_c''$ is the value of the complex Young modulus at a reference angular frequency $\omega_c$, chosen within the measured frequency band, $E_c'$ and $E_c''$ being the storage and loss moduli at $\omega=\omega_c$, respectively.
Although mathematically equivalent to Eq.~\eqref{eq:voigt}, this form replaces $(E_0, \tau)$ with the storage and loss moduli expressed at a frequency within the measured band. $E_0$ is the modulus in the quasi-static limit, which sits outside the limits of the measured band: recovering it requires extrapolating the power-law down to $\omega = 0$, so any uncertainty in $n$ propagates to it. In contrast, $(E_c', E_c'')$ are directly constrained by the measurements and are therefore largely insensitive to the estimation of $n$.

To address the second difficulty, we build an objective function that does not require an explicit assignment between measured and predicted points. We seek to recover the three parameters $\bm{\theta} = [E_c',\,E_c'',\,n]$ by minimizing a distance between the set of predicted dispersion points $\{(\omega_i,k_i(\bm{\theta}))\}$, $i=1,\ldots, N_i$, obtained by solving the forward problem at parameters $\bm{\theta}$, and the set of measured dispersion points $\{(\omega_j,\hat{k}_j,)\}$, $j=1,\ldots, N_j$, extracted with ESPRIT, with $N_i\neq N_j$ in general. Predicted and observed sets both include forward and backward propagating branches, treated as independent modes.  For a predicted point $i$ and an observed point $j$, we define the normalized distance
\begin{equation}
    d_{ij}(\bm{\theta}) = \sqrt{
    \left(\frac{\big|k_i(\bm{\theta}) - \hat{k}_j\big|}{\sigma_k}\right)^{2}
    + \left(\frac{f_i - \hat{f}_j}{\sigma_f}\right)^{2}}
    \label{eq:pairwise_distance}
\end{equation}
where $\sigma_k$ and $\sigma_f$ are fixed scale tolerance parameters that drive the range at which the objective function is sensitive to a predicted branch.

We associate each observed data point $j$ with a soft matching score, which is expressed by the log-sum-exp function of the normalized distance between observed and predicted dispersion curves:
\begin{equation}
    m_j(\bm\theta) = \log\Bigg(\frac{1}{N_i}\sum_{i=1}^{N_i}
    \exp\big(-d_{ij}(\bm\theta)\big)\Bigg)
    \label{eq:soft_match}
\end{equation}
 The matching score $m_j(\bm\theta)$ is a smooth function of $\bm\theta$, for which only the predicted points located near an observed point contribute significantly. Since the cost is built from one score per observed point, each measured point only needs to lie close to one predicted branch; predicted branches far from any measurement contribute negligibly, so the model is not penalized for predicting unexcited or evanescent solutions.

In order to discard points likely to result from noise overfitting in the ESTER mode-count selection (recall Section~\ref{subsec:method:esprit}), a thresholding weight is applied based on the RMS amplitude associated with each point (Equation~\eqref{rms}), expressed in decibels: $\hat{A}_j = 20\log_{10}(\hat{a}_j^{\text{RMS}})$:
\begin{equation}
    w_j =
    \begin{cases}
        1, & \text{if}~\hat{A}_j > A_{\text{noise}} \\[4pt]
        0, & \text{otherwise}
    \end{cases}
    \label{eq:weight}
\end{equation}
with $A_{\text{noise}}$ a threshold set around the expected noise floor level. 
The optimal set of parameters $\hat{\bm\theta}$ is then recovered by maximizing the weighted sum of matching scores, \textit{i.e.} finding:
\begin{equation}
    \hat{\bm\theta} = \arg\min\big(C(\bm\theta)\big),~~~~  C(\bm\theta)= -\sum_{j=1}^{N_j} w_j\, m_j(\bm\theta)
    \label{eq:objective}
\end{equation}
The optimization is carried out using the interior-point algorithm (\texttt{fmincon}, MATLAB Optimization Toolbox). This optimization method allows bounds to be defined for each parameter, chosen to be within an order of magnitude of the expected solution, with the initial value selected as the midpoint. We also systematically verified that the solutions remained robust to random variations in the initial condition, within the specified limits, in order to avoid local minima.

\section{\label{sec:experimental methods} Experimental methods}

\subsection{\label{subsec:experimental methods:sample fabrication} Sample fabrication}

Two homogeneous tube samples were fabricated from Ecoflex\textsuperscript{\textregistered} (Smooth-On) soft silicone elastomers: Ecoflex 00-30 (E30) and Ecoflex Sorta-Clear (SC12), the latter being stiffer than the former. The quasi-incompressible and viscoelastic nature of these materials makes them well suited to mimic soft biological tissues such as arterial walls. Furthermore, the materials have been extensively used to fabricate other  waveguides geometries for elastic waves, such as membranes or cylinders \cite{lanoyDiracConesChiral2020,delorySoftElastomersPlayground2022,chantelotWavePropagationModel2026,chantelotGuidedElasticWaves2026,croquetteGuidedElasticWaves2026,zhuYoungsModulusEstimation2025}. Ecoflex is a two-part platinum-catalyzed silicone elastomer; a liquid base and a liquid curing agent are mixed in equal parts and poured into a custom 3D-printed mold to obtain the desired tube geometry. The inner radius and wall thickness were measured as $r = 4.0 \pm 0.1~\mathrm{mm}$ and $h = 2.0 \pm 0.1~\mathrm{mm}$, respectively. This corresponds to approximately twice the dimensions of a typical human carotid artery, a necessary up-scaling to ensure sample integrity when removing it from the mold. Sample length was $\ell\approx30$~cm. Colored particles were embedded in the silicone matrix prior to casting, creating a speckle pattern on the tube surface for optical motion tracking.

A third tube sample was fabricated with graded mechanical properties along its axis: the mold was first half-filled with E30, then the remaining half was filled with SC12 before the first half had gelled, allowing the two materials to interdiffuse at the interface rather than form a sharp boundary. As for the homogeneous samples, particles were embedded in the silicone matrix but here, different colors were used for E30 and SC12. While this color distinction plays no role in the measurements themselves, it simply allows to confirm visually that the fabrication process, despite being difficult to control, indeed results in a continuous transition zone spanning almost the entire length of the tube. 

\subsection{\label{subsec:experimental methods:optical}  Optical wavefield acquisition and digital image correlation}

\begin{figure}[b]
    \centering
    \includegraphics[width=0.65\linewidth]{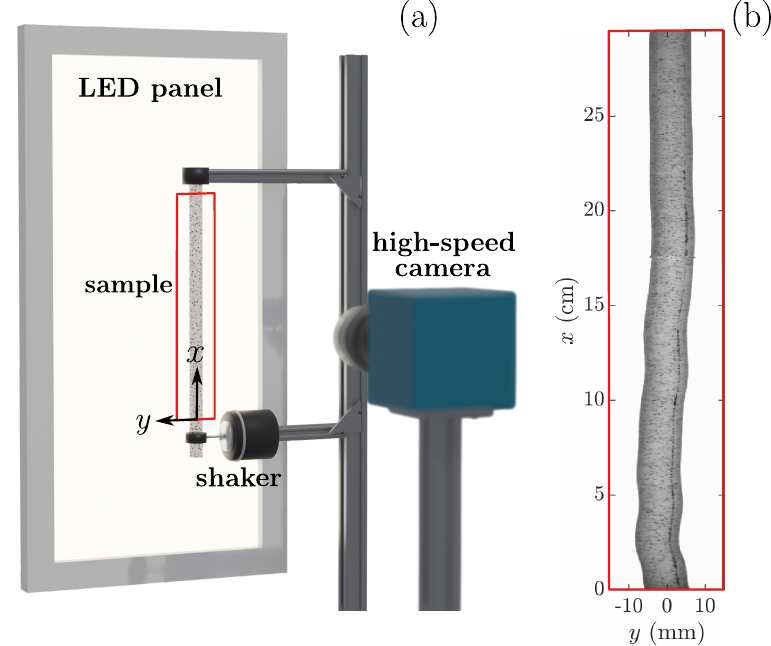}
    \caption{(a) Schematic representation of the experimental configuration. (b) Single snapshot of the homogeneous E30 sample.}
    \label{fig:exp_setup}
\end{figure}
Samples were suspended vertically in front of an LED backlight and filmed with a high-speed camera (Cube4, MotionBLITZ\textsuperscript{\textregistered}) at 600~frames per second over a full 30-second acquisition period. With the sample clamped at the top, a shaker (minishaker type 4810, HBK\textsuperscript{\textregistered}) was attached at the bottom of the sample and oriented laterally, so as to preferentially excite flexural waves (see Figure~\ref{fig:exp_setup}). The shaker was attached after the sample has been suspended in order to minimize as far as possible any static stress other than that caused by its own weight. The shaker continuously emitted white Gaussian noise, low-pass filtered below 250~Hz to remain within the Nyquist frequency of the camera. White Gaussian noise was chosen to ensure broadband excitation across the entire frequency range under consideration, but also to emphasize that the method does not require a coherent source signal, since it relies entirely on spatial correlation.

A conventional speckle-tracking digital image correlation (DIC) procedure \cite{guizar-sicairosEfficientSubpixelImage2008} was employed to extract the local two-dimensional inter-frame displacement field from pairs of successive frames. In short, each frame (around $160 \times 1550$~pixels) was subdivided into $40 \times 20$~pixel windows with 75\% overlap. For each window, the local two-dimensional cross-correlation was computed as the inverse Fourier transform of the product of the spatial Fourier transform of the window in the current frame and the complex conjugate of the corresponding window's Fourier transform in the previous frame. The local displacement was first estimated as the position of the correlation peak, yielding a coarse, pixel-locked approximation, which was subsequently refined to $1/10^{\text{th}}$  of a pixel using a direct Fourier transform interpolation, thereby enabling the detection of displacement amplitudes down to approximately $19$~$\mu$m.

\begin{figure}[htbp]
	\centering
	\includegraphics[width=0.5\linewidth]{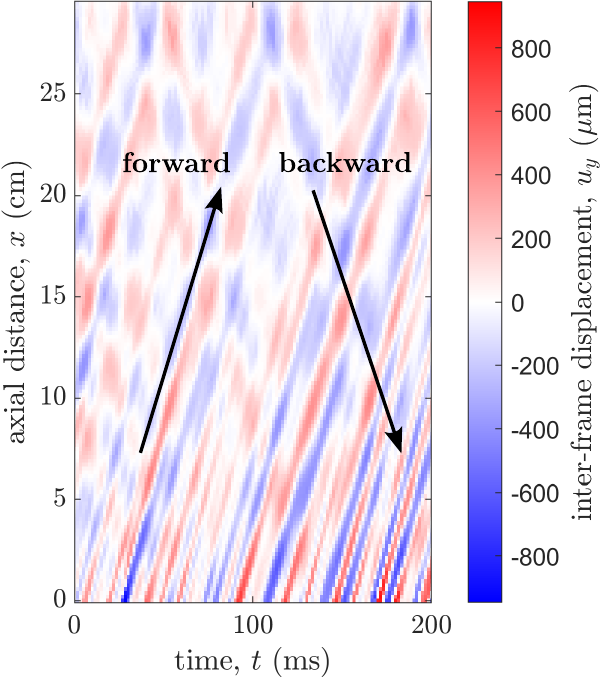}
	\caption{A 200-ms sample of the spatio-temporal signal obtained using digital image correlation on homogeneous E30 sample.}
	\label{fig:Bscan}
\end{figure}

Although the method extracts both horizontal and vertical displacement fields, only the transverse displacement $u_y$, which is orthogonal to the waveguide axis and predominant in flexural motion, is kept for further analysis. Since the fundamental flexural mode $F(1,1)$ displacement is uniform across the tube cross-section, $u_y(x,y,t)$ is averaged over the transverse dimension $y$ to recover a spatio-temporal signal $u_y(x,t)$,  taking the form of a B-scan  with $N=155$ spatial positions spaced by $\Delta x = 1.89$~mm. The total field shown in Figure \ref{fig:Bscan} results from the superposition of the noise signal excitation emitted by the shaker at $x=0$ and its reflection at the clamped end $x=L$. 

\FloatBarrier

\section{\label{sec:results} Results}

This section is split into two parts. In the first part, the method is employed to recover the complex Young modulus of E30 and SC12 materials by making use of the full axial length. This does not aim at demonstrating the super-resolution property of the method but only to provide global reference values. In the second part, the method is applied to small tube axial segments to test the capability of the method to recover the spatial dependence. The local method is tested  on both homogeneous and on the graded samples.

\subsection{\label{subsec:homogeneous} Global viscoelastic properties}

The total 30-s long signal is first split into 500~ms duration windows, treating them as $M=60$ separated measurements. A Fast Fourier transform is applied to each window to convert the data from the time to the frequency domain. For each frequency, the stacked Hankel matrix is formed with the $M$ windows and is used as input for ESPRIT with $L=76 \simeq N/2$ and testing ESTER criterion up to $P=2$ with $\gamma = 0.2$ (recall Section~\ref{subsec:method:esprit}). The model based inverse problem is obtained with $\sigma_f=5$~Hz and $\sigma_k=5$~rad/m (recall Section~\ref{subsec:method:inverse})

\begin{figure*}[htbp]
    \centering
    \includegraphics[width=0.99\linewidth]{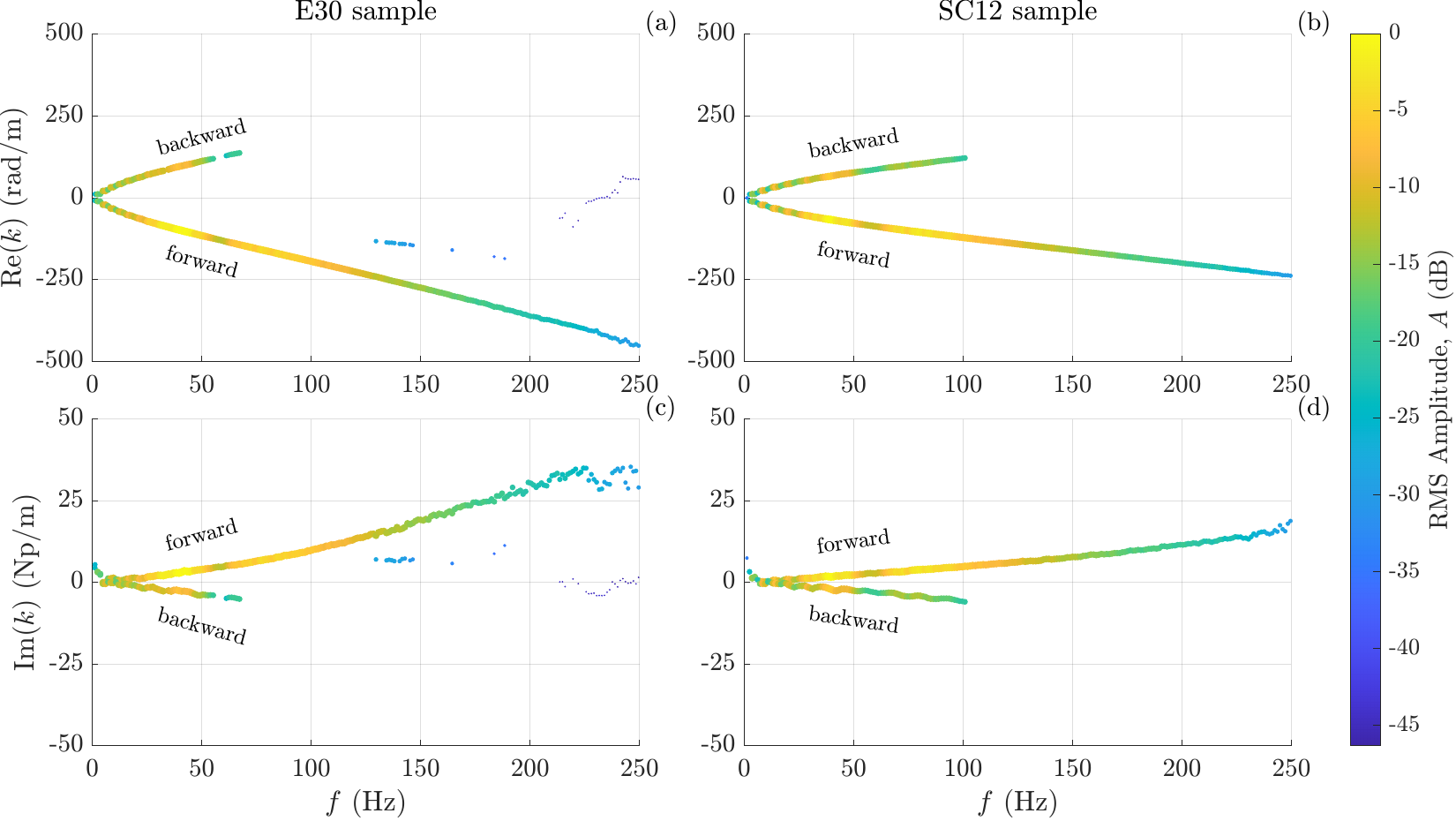}
    \caption{Wavenumbers extracted with ESPRIT for the E30 (a,c) and SC12 (b,d) homogeneous samples. (a,b) Real part of the wavenumber, $\mathrm{Re}(k)$, (c,d) Imaginary part, $\mathrm{Im}(k)$. Because of the $e^{i\omega t}$ convention, the sign of the imaginary branch is reversed relative to that of the real part. In all panels, marker size and color both encode the RMS amplitude $A$ in dB. (Sec.~\ref{subsec:method:esprit}) }
    \label{fig:ESPRIT_homogeneous}
\end{figure*}

Figure~\ref{fig:ESPRIT_homogeneous} shows the complex dispersion relations recovered using ESPRIT for the homogeneous E30 and SC12 samples. The real part of the wavenumber splits into two branches of opposite sign, the negative branch corresponding to forward propagation and the positive branch to backward propagation of the fundamental flexural mode. The amplitude associated with each modal solution (see Equation~\eqref{rms}) generally decreases with frequency as attenuation plays a greater role. The backward-propagating branch is further affected: having traveled a greater distance, it is attenuated more strongly, to the point where it blends into the background noise and is no longer detected by ESPRIT. In addition to the clearly visible flexural mode, ESPRIT detects a few other points, in particular for the E30 sample. Comparison with the model (see section~\ref{sec:model}) shows that the line with markedly lower amplitude at 140~Hz aligns with the torsional mode that may be weakly excited as a result of minor imperfections in the experimental setup. The other very-low-amplitude points beyond $200$~Hz result from slight noise overfitting. In any case, these extra solutions are not accounted for in the model-based inversion, since they are set aside by applying an amplitude threshold at $A_{\text{noise}} = -25$ dB (recall Section \ref{subsec:method:inverse}) relative to max amplitude.

\begin{figure}[htbp]
    \centering
    \includegraphics[width=0.7\linewidth]{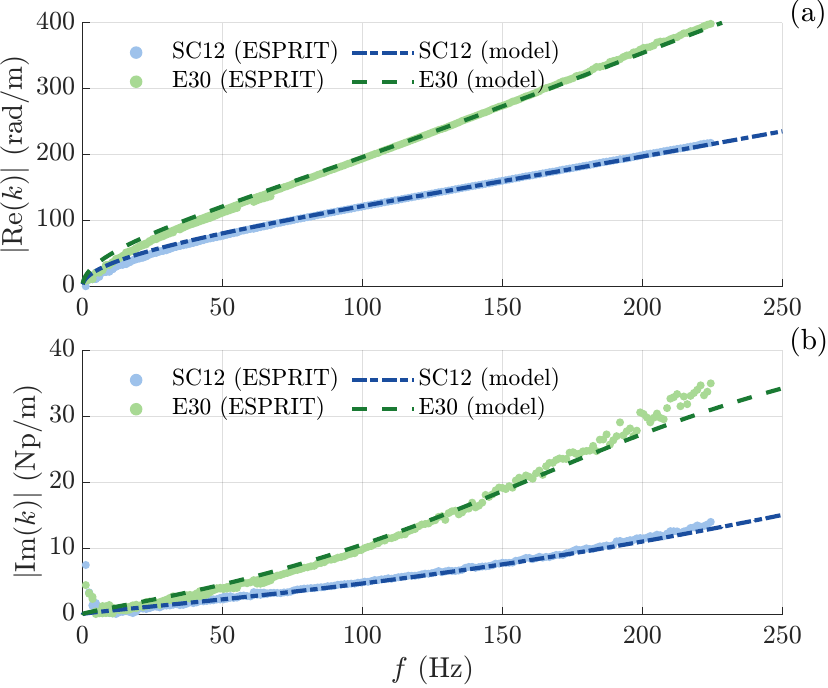}
    \caption{Comparison between the wavenumbers extracted with ESPRIT (markers) and the fractional Kelvin--Voigt model recovered from the inverse problem (dashed lines), for the E30 and SC12 homogeneous tubes. (a) Magnitude of the real part, $|\mathrm{Re}(k)|$; (b) magnitude of the imaginary part, $|\mathrm{Im}(k)|$.}
    \label{fig:fit_homogeneous}
\end{figure}

\begin{figure}[htbp]
	\centering
	\includegraphics[width=0.6\linewidth]{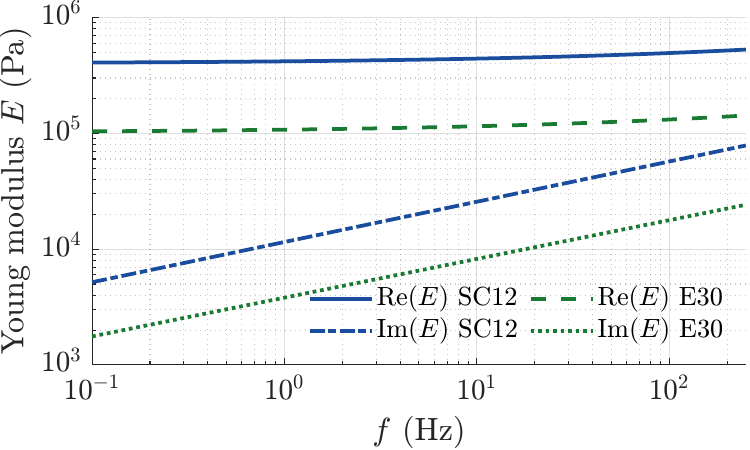}
	\caption{Frequency-dependent complex Young's modulus obtained from the inverse problem for homogeneous E30 and SC12 samples.}
	\label{fig:young_homogeneous}
\end{figure}

Figure \ref{fig:fit_homogeneous} compares the dispersion relation obtained from model-based inversion with that obtained using ESPRIT. For better readability, this graph shows only the magnitudes of the real and imaginary parts; the forward and backward waves are therefore superimposed in this representation, but the optimization does take into account both positive and negative solutions separately. A good fit is obtained for both materials and for both the real and imaginary parts of the wavenumber. The corresponding parameters of the fractional Kelvin–Voigt model at the reference frequency $f_c$ are presented in Table \ref{tab:params}, and the complex Young’s moduli as a function of frequency are shown in Figure~\ref{fig:young_homogeneous}.
The reference frequency value $f_c = 100$~Hz is chosen within the measured frequency band to avoid an ill-conditioned problem; the reported frequency dependent complex Young moduli are not sensitive to the specific choice of $f_c$ within this band.

\begin{table}[H]
	\centering
	\setlength{\tabcolsep}{6pt}
	\begin{tabular}{ccccccc}
		\hline\hline
		& \multicolumn{3}{c}{Fitted} & \multicolumn{3}{c}{Deduced} \\ \cline{2-4} \cline{5-7}
		& \begin{tabular}[c]{@{}c@{}} $\hat{E}_c'$\\ (kPa)\end{tabular}
		& \begin{tabular}[c]{@{}c@{}} $\hat{E}_c''$\\ (kPa)\end{tabular}
		& \begin{tabular}[c]{@{}c@{}} $\hat{n}$ \\    ~~ \end{tabular}
		& \begin{tabular}[c]{@{}c@{}} $\tan\delta_c$ \\   ~~ \end{tabular}
		& \begin{tabular}[c]{@{}c@{}} $\hat{E}_0$\\ (kPa)\end{tabular}
		& \begin{tabular}[c]{@{}c@{}} $\hat{\tau}$\\ ($\mu$s)\end{tabular} \\ \hline
		\begin{tabular}[c]{@{}c@{}}Ecoflex 00-30\\ (E30)\end{tabular} & 131.5 & 17.8 & 0.335 & 0.135 & 100.8 & 69.9 \\
		\begin{tabular}[c]{@{}c@{}}Sorta-Clear\\ (SC12)\end{tabular} & 493.0 & 57.1 & 0.347 & 0.116 & 399.0 & 39.0 \\ \hline\hline
	\end{tabular}
	\caption{Viscoelastic parameters fitted at the reference frequency $f_c = 100$~Hz, and the corresponding deduced loss tangent $tan\delta_c=\hat{E}_c''/\hat{E}_c'$, static modulus $\hat{E}_0$ and relaxation time $\hat{\tau}$.}
	\label{tab:params}
\end{table}

As expected, SC12 is stiffer than E30, with a ratio of 3.7 for the real part of the Young modulus at $f_c$. Since the imaginary part of SC12's Young modulus is only approximately 3.2 times greater than that of E30, the loss tangent at the reference frequency ($\tan\delta_c = \hat{E}_c''/\hat{E}_c'$) is lower for SC12 (0.116 versus 0.135), confirming that E30 is indeed lossier than SC12. Furthermore, both materials are measured with a very similar frequency dependence, following a power-law with an exponent of approximately $n=0.34$ for both (\textit{i.e.} the slope of the imaginary part of the Young modulus on a log-log scale in Figure~\ref{fig:young_homogeneous}), which suggests a likely common attenuation mechanism.

\FloatBarrier
\subsection{\label{subsec:graded} Local viscoelastic properties}

The same signal processing is now applied to local axial segments rather than on the full length of the tube. The axial dimension $x$ is divided into 6 segments of approximately $8$~cm with 50\% overlap. ESPRIT is applied independently on each of them. The size of 8~cm for the segment length (\textit{i.e.} roughly one wavelength at 50~Hz) has been chosen empirically as an observed limit beyond which the recovery of the imaginary starts to degrade too severely. 

\begin{figure*}[b]
    \centering
    \includegraphics[width=0.95\linewidth]{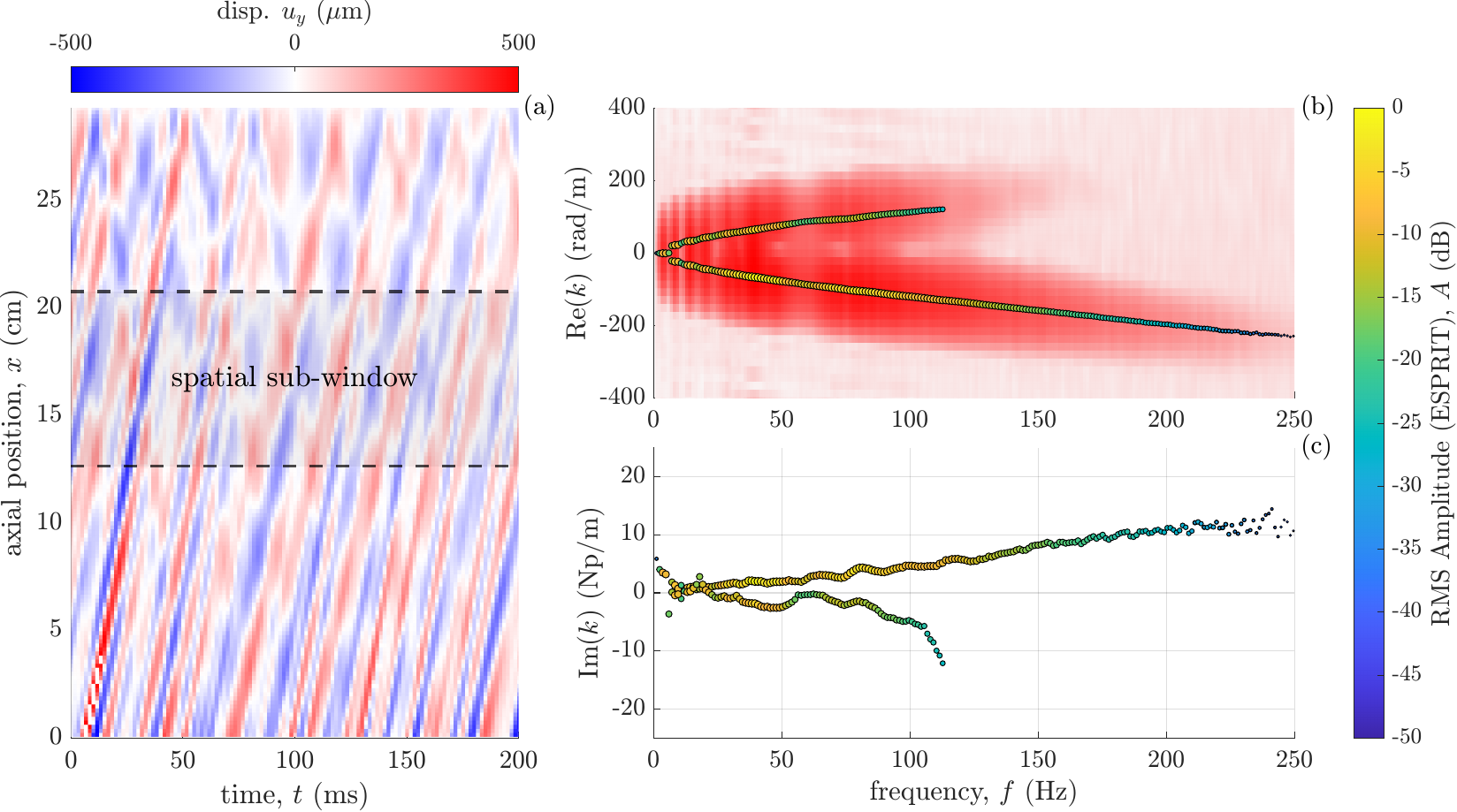}
    \caption{(a) B-scan of the inter-frame displacement $u_y(x,t)$ measured on the SC12 homogeneous sample; the region between the dashed lines indicates the axial segment (of approximately one wavelength, recall
Sec.~\ref{subsec:method:esprit}) used for the local dispersion estimation shown in (b,c). (b,c) Complex wavenumber extracted with ESPRIT from this segment: (b) real part, $\mathrm{Re}(k)$; (c) imaginary part, $\mathrm{Im}(k)$. The red background represents the raw magnitude of the two-dimensional Fourier transform of the same segment with a Hamming window applied, shown on the same dynamic range ($-50$ to $0$~dB) for comparison.}
    \label{fig:high_resolution}
\end{figure*}
Figure~\ref{fig:high_resolution} illustrates ESPRIT applied to the fourth spatial segment of the SC12 homogeneous tube. It should be noted that the reported wavenumber corresponds to an \emph{averaged} effective wavenumber over the entire width of the segment. The result is compared with the magnitude of the standard 2D Fourier transform over the same segment. While blind to the imaginary part of the wavenumber by construction, the magnitude of the Fourier transform strongly suffers from the limited spatial resolution, resulting in very large and partially overlapping main lobes. Conversely, ESPRIT still manage to recover both real and imaginary with a good precision, despite the limited aperture: this illustrates the \textit{super-resolution} property of the method. The cost of the limited segment size appears to be a degraded accuracy, especially for the imaginary part of backward wave, compared to full domain processing (recall Figure~\ref{fig:ESPRIT_homogeneous}(d)). 

The same segmented signal processing is applied for both homogeneous samples as well as for the graded one. The exponent parameter $n$ was held constant at $n=0.34$, and only $E_c'$ and $E_c''$ were adjusted when solving the inverse problem using locally observed dispersion curves.

\begin{figure}[htbp]
    \centering
    \includegraphics[width=0.7\linewidth]{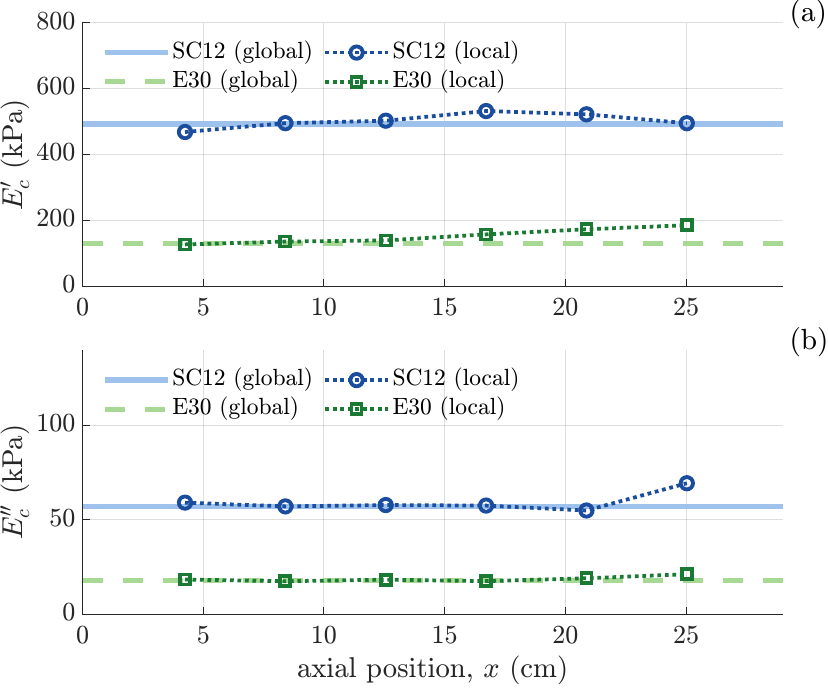}
    \caption{Local viscoelastic estimates for the two homogeneous samples, E30 and SC12, as a function of axial position along the tube: (a) storage modulus $E_c'$, (b) loss modulus $E_c''$ at reference frequency $f_c=100$~Hz. Markers connected by dotted lines show estimates obtained independently on each spatial window; horizontal lines show the corresponding estimate obtained from the full tube length. }
    \label{fig:homoegeneous_local}
\end{figure}
Figure \ref{fig:homoegeneous_local} displays the recovered storage modulus $E_c'$ and loss modulus $E_c''$ at reference frequency $f_c$, estimated locally and globally, for the homogeneous samples. The $x$-axis position of each marker indicates the center of the corresponding segment along the axial direction. Local estimates are consistent with the global value for both materials and both quantities, indicating that the method recovers a correct estimate of viscoelastic properties even from a reduced spatial aperture. A small, systematic increase with axial position is nonetheless visible in both storage and loss modulus for E30, while SC12 remains relatively close to its global value throughout.
\begin{figure}[htbp]
    \centering
    \includegraphics[width=0.7\linewidth]{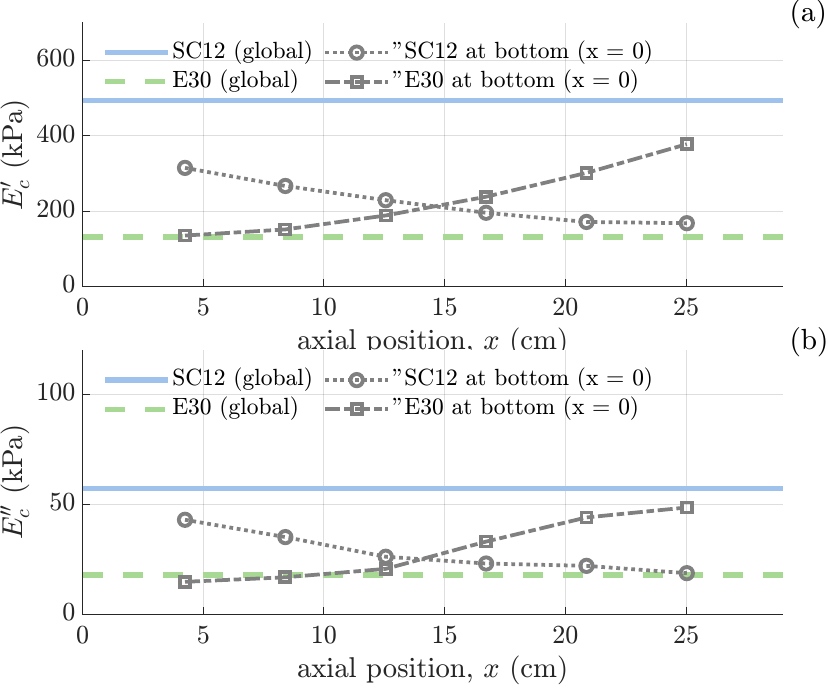}
    \caption{Local viscoelastic estimates for the axially graded sample, measured in its original orientation (E30 at bottom) and after flipping the tube end-to-end (SC12 at bottom): (a) storage modulus $E_c'$, (b) loss modulus $E_c''$ at reference frequency $f_c=100$~Hz. Horizontal lines recall the homogeneous global values of E30 and SC12 as reference bounds.}
    \label{fig:graded_local}
\end{figure}
Figure \ref{fig:graded_local} displays the recovered storage modulus $E_c'$ and loss modulus $E_c''$ of the graded sample with in both possible orientation (with the stiffer end either at the top or at the bottom). In both cases, the local estimates transition smoothly between the two bounds, and reversing the tube orientation reverses the recovered spatial profile, approximately mirroring it about the midpoint. This confirms that the recovered spatial variation reflects the physical gradient of the tube rather than an artifact of scan direction or windowing. As with the homogeneous tubes, a small and systematic asymmetry is observed. It corresponds to a slight and consistent increase with axial position from bottom to top which will be further discussed in section \ref{sec:discussion}. 

It was decided not to present error bars here. In fact, an honest, comprehensive, and truly informative quantification of the measurement error would have required not only repeating the fabrication and measurement of the samples, but also quantifying the uncertainty associated with the inverse problem using a sensitivity analysis, as well as quantifying and accounting for the impact of uncertainties on the parameters considered known (radius, thickness, fluid properties, density). Instead, we propose a theoretical analysis of the fundamental limitations associated with the local measurement of complex dispersion curves, which we consider to be more relevant and informative in this particular case.

\section{\label{sec:uncertainty} Quantifying fundamental precision limits}

The Cram\'er--Rao bound (CRB) is a lower bound on the variance of any unbiased estimator of a parameter recovered from a noisy measurement: it is a benchmark attached to the measurement itself, independent of the estimator one eventually builds, and no unbiased estimator can outperform it. It tells us, even before choosing an algorithm, that the precision of the parameter estimates we seek is naturally limited by the amount of information that can be extracted from a finite, noisy signal.

We begin with the simplest case, a single lossless wave (\textit{i.e.} the only unknown then is the real wavenumber $k_1'$), whose scaling laws set the reference that the attenuated and realistic cases inherit. The CRB constrains the variance of the estimated $\hat{k}_1'$ and is given by (see Appendix~\ref{app:crb} for details):

\begin{equation}
\label{eq:crb_lossless1}
    \text{var}(\hat{k}_1') \geq \text{CRB}(k_1') = \dfrac{\sigma^2}{2|a_1|^2\displaystyle\sum_n x_n^2} \quad ,
\end{equation}
where $\sigma^2$ is the variance of the additive noise and $a_1$ is the complex amplitude (recall Equation~\ref{eq:signal_model}). This can be written as a function of the relative standard deviation and further simplified as:
\begin{equation}\label{eq:crb_lossless2}
    \frac{\mathrm{std}(\hat k_1')}{k_1'} \geq \frac{\sqrt{\mathrm{CRB}(k_1')}}{k_1'}
    \;\gtrsim\; \,\frac{1}{\mathrm{SNR}\cdot k_1' L \cdot\sqrt{N}}.
\end{equation}
where $L=N\Delta x$ is the total axial segment length and $\mathrm{SNR}=|a_1|/\sigma$. Two characteristics are common to everything that follows. First, the bound obviously vanishes as $\sigma\to0$: in the noiseless limit the wavenumber can in principle be recovered exactly, regardless of the values of the other parameters. Second, the limit depends only on the relative aperture $k_1'L$, the SNR, and the number of measurement points $N$. To remain under a target relative error $\beta = \mathrm{std}(\hat k_1')/k_1'$, the signal-to-noise ratio must satisfy the condition $\text{SNR}\gtrsim 1/(\beta\sqrt{N}k_1'L)$. In other words, a twofold deterioration in the SNR can be compensated for either by an aperture $L$ that is twice as long, by a wavelength $\lambda = 2\pi/k_1'$ that is halved, or by a number of measurement points $N$ that is quadrupled.

Now reintroducing the imaginary part, the same wave carries two parameters and acquires a decay component, $e^{ik_1' x_n}e^{-k_1'' x_n}$. For a single complex wavenumber and a uniform array $x_n=n\Delta x$, the real and imaginary parts fully decouple (Appendix~\ref{app:crb}), meaning that varying the value of either the real or imaginary part will not affect the accuracy limit for the estimation of the other. Moreover, both share a single absolute bound,
\begin{equation}\label{eq:crb_lossy}
    \mathrm{CRB}(k_1') = \mathrm{CRB}(k_1'')
    = \frac{\sigma^2}{2|a_1|^2\sum_{n=1}^N x_n^2\,e^{-2k_1'' x_n}}.
\end{equation}
The expression is very similar to the lossless bound  Equation~\ref{eq:crb_lossless1}), the only difference being that the effective aperture is also reduced by attenuation. The weight $e^{-2k_1'' x_n}$ suppresses the information from the most distant measurement points. Beyond roughly one attenuation length $1/k_1''$,  the signal falls into the noise floor and no longer contributes significantly to information. By its very nature, attenuation reduces the useful information available for its own estimation.

\begin{figure}[t]
    \centering
    \includegraphics[width=0.7\linewidth]{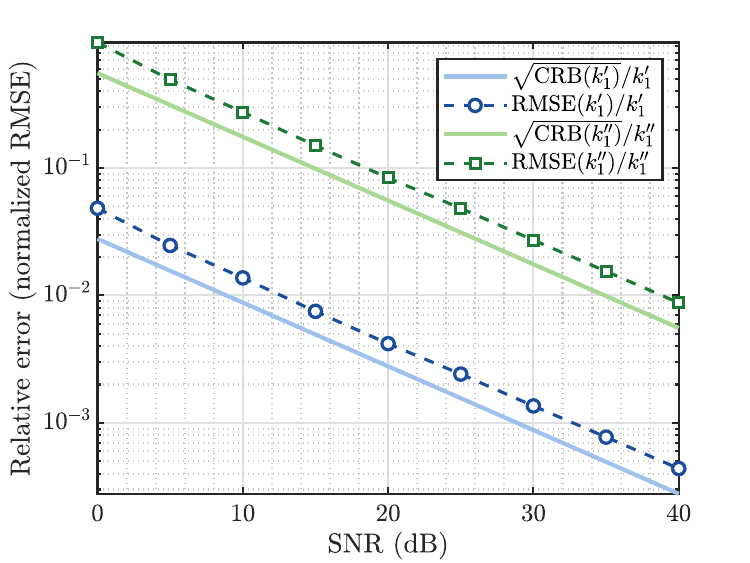}
    \caption{ relative Root-Mean-Square Error (RMSE) of ESPRIT estimator compared to Cram\'er-Rao bound. RMSE is computed using Monte Carlo simulation on 1000~realizations of a monochromatic damped wave ($k_1'/k_1'' = 20$, $N=100$) over a constant relative segment size equal to one wavelength, $L=\lambda$, with different level of noise.}
    \label{fig:crb}
\end{figure}

Figure~\ref{fig:crb} plots the relative-error bound (Equation~\ref{eq:crb_lossy}) against varying signal-to-noise ratio, together with the empirical relative errors of the ESPRIT estimates obtained by Monte~Carlo simulation on a synthetic damped monochromatic field.
ESPRIT stays close to the bound across the whole range of SNR, meaning it almost reaches the maximum theoretical accuracy. Although the real and imaginary parts share the same \emph{absolute} bound \eqref{eq:crb_lossy}, their \emph{relative} bounds differ by the ratio $k_1'/k_1''$:
\begin{equation}\label{eq:eta_gap}
    \frac{\mathrm{std}(\hat k_1'')}{k_1''}
    = \frac{k_1'}{k_1''}\,\frac{\mathrm{std}(\hat k_1')}{k_1'}.
\end{equation}
This offset constitutes a central consideration for attenuation recovery  since  $k_1'/k_1''$ is usually relatively large in most practical applications, \textit{i.e.} the characteristic attenuation length is much larger than the wavelength. The implication can be clearly observed in Figure~\ref{fig:crb}, for instance with $k_1'/k_1''=20$, The relative Root-Mean-Square Error (RMSE) from Monte-Carlo simulation yields a 4.7\% error on the real part of the wavenumber at SNR = 0~dB, associated with a 95.2\% error on the imaginary part. It can also be thought of as a vertical offset: estimating the attenuation to the same relative precision as the real wavenumber demands a factor $k_1'/k_1''$ in signal-to-noise ratio, an extra $20\log_{10}\big(k_1'/k_1''\big)$~dB of headroom, or 26 dB in this specific case. The imaginary part is thus harder to estimate than the real part with the same relative accuracy; the CRB sets the SNR and aperture conditions a measurement needs to meet in order to be able to evaluate attenuation reliably.

Although this analysis provides useful information on the limits of achievable accuracy, it is based on identified simplifications. The derivation here assumes the amplitude is known, which is not the case in the experiment and rarely the case in practice. Leaving amplitude as a free parameter would only loosen the bound, so the CRB above is optimistic and is best read as the scaling law governing wavenumber estimation rather than as a tight numerical floor. Moreover, this analysis treats a single wave, whereas a real field may carry multiple modes together with their counter-propagating reflections; the single-wave bounds treated here remain a reasonable approximation only when considering relatively well-resolved modes. The exact impact of an unknown amplitude and the limiting case where two solutions are very close to one another therefore remain to be investigated in future works.

\section{\label{sec:discussion}Discussion}

The validity of the estimates is first evaluated on homogeneous samples of E30 and SC12 over the entire available spatial domain. In the absence of an independent measurement on our samples, our estimates are compared with values reported in previous works. Fractional Kelvin--Voigt parameters have been reported from guided-wave inversion on thin strips and from a conventional rheometer for both materials~\cite{deloryViscoelasticDynamicsSoft2024,chantelotGuidedElasticWaves2026,croquetteGuidedElasticWaves2026,croquetteGuidedElasticWaves2026}, although only one work covers SC12. These values are gathered in Table~\ref{tab:literature}, expressed as the complex Young modulus at $f_c = 100$~Hz, a frequency covered by both methods. For both of these materials, our estimates of the dynamic storage and loss moduli sit reasonably close to the reported values. The quasi-static Young moduli have also been independently evaluated from tensile tests in some of these works~\cite{deloryViscoelasticDynamicsSoft2024,chantelotGuidedElasticWaves2026,croquetteGuidedElasticWaves2026}. By comparison, the quasi-static moduli extrapolated from our model, $E_0 = 100.8$~kPa for E30 and $399.0$~kPa for SC12, exceed the values obtained from tensile tests, namely $67$--$70$~kPa for E30~\cite{deloryViscoelasticDynamicsSoft2024,chantelotGuidedElasticWaves2026,croquetteGuidedElasticWaves2026} and $381$~kPa for SC12~\cite{croquetteGuidedElasticWaves2026}. This discrepancy is to be expected. First, the quasi-static value is not directly measured but has to be extrapolated from higher frequencies and is therefore very sensitive to the estimation of the exponent $n$, itself difficult to recover precisely over a limited frequency range. Indeed, a more accurate quantification of the exponent would require covering a broader spectrum~\cite{liuUltraWideband05162014}. Second, the measured dispersion combines the intrinsic dispersion of flexural waves with the material dispersion due to viscoelasticity. Any effect that modifies the dispersion but is not accounted for in the forward model (e.g., static prestress or geometric uncertainties) is compensated by the viscoelastic parameters in the inverse problem, thereby introducing a bias for their estimation.

\newcommand{\tabref}[2]{#1 et al.\cite{#2}}
\begin{table}[b]
	\centering
	\setlength{\tabcolsep}{6pt}
	\begin{tabular}{lllc}
		\hline\hline
		Material & Method & Reference & $E_c$ (kPa) \\ \hline
		E30  & Guided waves & \tabref{Lanoy}{lanoyDiracConesChiral2020}          & $115.3 + 21.3i$ \\
		&              & \tabref{Croquette}{croquetteGuidedElasticWaves2026} & $140.5 + 23.6i$ \\
		& Rheometer   & \tabref{Delory}{deloryViscoelasticDynamicsSoft2024} & $105.5 + 20.1i$ \\
		&              & \tabref{Chantelot}{chantelotGuidedElasticWaves2026} & $114.7 + 19.8i$ \\
		&              & \tabref{Croquette}{croquetteGuidedElasticWaves2026} & $\approx 110 + 20i$ \\
		& This work    &                                                     & $131.5 + 17.8i$ \\ \hline
		SC12 & Guided waves & \tabref{Croquette}{croquetteGuidedElasticWaves2026} & $671.8 + 36.7i$ \\
		& Rheometer    & \tabref{Croquette}{croquetteGuidedElasticWaves2026} & $\approx 380 + 40i$ \\
		& This work    &                                                     & $493.0 + 57.1i$ \\ \hline\hline
	\end{tabular}
	\caption{Complex Young modulus $E_c$ at $f_c = 100$~Hz for E30 and SC12. Literature values are computed from the reported fractional Kelvin--Voigt parameters, with $E = 3\mu$ when shear moduli are reported; values marked $\approx$ are read from graphs.}
	\label{tab:literature}
\end{table}

In particular, our model does not account for acousto-elastic effects, which are known to significantly affect wave propagation in soft elastomers~\cite{deloryViscoelasticDynamicsSoft2024,croquetteGuidedElasticWaves2026,chantelotWavePropagationModel2026}. While no prestress was intentionally applied in the experimental setup, soft materials are nevertheless stressed by their own weight, which increases their apparent stiffness. Two observations suggest that this effect, although weak, is visible in our measurements. First, the dispersion relation obtained with ESPRIT is very slightly, but consistently, less dispersive than the fitted model at low frequencies (see Figure~\ref{fig:fit_homogeneous}(a)), which corresponds to the expected effect of an axial tension. Second, local estimates of the modulus tend to be higher on the side where the tube is suspended, regardless of which tube is considered or its orientation (recall Figures~\ref{fig:homoegeneous_local} and~\ref{fig:graded_local}). This is, in fact, the part subjected to the highest tension, since it supports the entire weight of the tube below it. Although identifying this effect was not the aim of this study, it can be detected only because the dispersion relation is measured independently of the model, and with sufficient accuracy. 

It is empirically well known that estimating the imaginary part of the wavenumber is more difficult than estimating the real part, particularly when working with a limited spatial domain. The CRB analysis demonstrates that this is, in fact, a fundamental limit related to the finite amount of information contained in the signal. This is also not specific to ESPRIT, which is a nearly efficient estimator in this respect. More precisely, the CRB reveals that (i) the accuracy of the recovered wavenumber is fundamentally limited by a combination of the signal-to-noise ratio (SNR) and the ratio of the wavelength to the segment length, and (ii) the relative accuracy of the imaginary part is necessarily lower than that of the real part by a factor $k'/k''$ (see Equation~\eqref{eq:eta_gap}). Although useful for understanding these limitations, the CRB is itself an idealization that assumes, among other things, additive white Gaussian noise; it therefore cannot be used directly to deduce a comprehensive measurement uncertainty.

Beyond that, the water-filled elastomer tube considered here remains a strong simplification of a real artery, and modeling arterial wave propagation is notoriously difficult. The arterial wall is anisotropic and layered~\cite{jiangComprehensiveCharacterizationNonlinear2026}, and the most adequate rheological model for soft living tissue is still debated~\cite{parkerConsensusRheologicalModels2019}. Moreover, the wall undergoes large deformations, it is prestressed by axial tension and by a blood pressure that varies over the cardiac cycle, and waves leak into the surrounding tissue as well. Any forward model is therefore necessarily approximate. In a single-step inversion, such as full-waveform inversion, a model mismatch is absorbed into the estimated parameters while the fit may still look convincing, leaving little or no way to detect it. The two-stage strategy adopted here does not remove this bias, but it makes it detectable. Since the complex wavenumbers are extracted without assuming any constitutive law or waveguide model, the measured dispersion relation constitutes a separate observable against which any candidate model can be tested, as illustrated by the gravity-induced prestress discussed above. 

Finally, the proposed approach is not, in principle, limited to optical measurements. ESPRIT makes no assumption about the imaging modality, it only requires a wavefield uniformly sampled along the propagation axis. Within these constraints, the same method would apply equally to displacement fields measured optically, by ultrafast ultrasound imaging or by magnetic resonance elastography. 

\section{\label{sec:conclusion}Conclusion}

This work proposes a two-stage method to recover local viscoelastic properties from 1D guided waves. It combines a high-resolution local estimation of the complex dispersion relation, including attenuation, performed independently of any constitutive or waveguide model, with a model-based inverse problem that matches the predicted dispersion relation to the measured one. The method was tested experimentally with flexural waves on water-filled soft tubes. The complex moduli obtained on homogeneous samples are consistent with values reported in the literature, and local estimates resolve spatial variations of the properties along the tube. Analysis of the Cramér–Rao bound further suggests that local viscoelastic characterization at a scale smaller than the wavelength is fundamentally limited, regardless of the inversion method used. Nevertheless, it remains possible and desirable to approach this limit. The ability of high-resolution wavenumber analysis to estimate local viscoelastic properties makes it particularly valuable for the characterization of arteries, where access to the acoustic displacement field is often restricted to small segments of the artery and where the features of interest, such as localized stiffening, are themselves often locally confined.

\section*{Acknowledgments}

We thank Daniel Kiefer for fruitful discussions regarding the use of the GEWtool code. This study was supported by the ITMO Cancer of Aviesan as part of the 2030 Cancer Control Strategy, using funds managed by Inserm under the Smart Imaging program.

\section*{Author Declaration}
The authors have no conflicts to disclose.

\section*{Data Availability}
The data that support the findings of this study are available from the corresponding author upon reasonable request.

\appendix

\section{Frequency-normalized fractional Kelvin--Voigt model}
\label{app:normalization}

The fractional Kelvin--Voigt rheological model can be rewritten in a frequency-normalized form with respect to an arbitrary frequency $f_c$. Introducing $\omega_c = 2\pi f_c$, Equation~\eqref{eq:voigt} can be written as:

\begin{equation} \label{eq:norm_c}
    E(\omega) = E_0 + E_0 (\omega_c\tau)^n\left(\dfrac{i\omega}{\omega_c}\right)^n.
\end{equation}

Expressing $E(\omega)$ at $\omega_c$, $E(\omega_c) = E_c' +i E_c''$ and making use of $i^n = \cos(n\pi/2) +i\sin(n\pi/2)$, we can identify:

\begin{subequations}
    \begin{align}
        E_c' &= E_0 +E_0(\omega_c\tau)^n\cos(n\pi/2) \label{eq:Ec'} \quad ,\\ 
        E_c'' &= E_0(\omega_c\tau)^n\sin(n\pi/2) \quad .\label{eq:Ec''} 
    \end{align}
\end{subequations}
Solving for $E_0$ and $\tau$ yields:

\begin{subequations}
    \begin{align}
        E_0 &= E_c' - E_c''\cot(n\pi/2)  \label{eq:E0} \quad , \\
        \tau &= \dfrac{1}{\omega_c}\left(\dfrac{E_c''}{E_0\sin(n\pi/2)}\right)^{\frac{1}{n}} \quad . \label{tau} 
    \end{align}
\end{subequations}

Reintroducing Equations \eqref{eq:E0} and \eqref{tau} into Equation \eqref{eq:norm_c}, we finally recover equation~\eqref{eq:voigt_normalized}:

\begin{equation}
    E(\omega) = \Big[E_c'-E_c''\cot(n\pi/2)\Big] + \dfrac{E_c''}{\sin(n\pi/2)}\left(i\dfrac{\omega}{\omega_c}\right)^n .
\end{equation}

\section{Derivation of the Cram\'er-Rao Bound} \label{app:crb}

\subsection{Fisher information matrix}

Starting from the noisy signal model:
\begin{equation}
    s_n = \bm{v}_n(\bm{\Psi})\bm{a}+b_n
\end{equation}
where $\bm{v}_n$ is the $n$-th row of the Vandermonde matrix and $\bm{\Psi}$ is the vector of unknown real parameters (in this case containing the real and imaginary parts of the wavenumbers), and $\bm{a}$ is the assumed known complex amplitude vector.

$b_n \sim \mathcal{CN}(0,\sigma^2)$ is a circularly-symmetric complex Gaussian noise variable with zero mean and total variance $\sigma^2$, split evenly and independently between real and imaginary parts. Since $\mu_n(\bm{\Psi}) = \bm{v}_n(\bm{\Psi})\bm{a}$ is purely deterministic, $s_n = \mu_n(\bm{\Psi}) + b_n$ is simply a deterministic shift of $b_n$, so
\begin{equation}
    s_n \sim \mathcal{CN}\big(\mu_n(\bm{\Psi}),\,\sigma^2\big) \quad,
\end{equation}
\textit{i.e.} the probability density of $s_n$ given $\bm{\Psi}$ is
\begin{equation}
    p(s_n;\bm{\Psi}) = \frac{1}{\pi\sigma^2}\,\exp\!\left(-\dfrac{\left|s_n-\mu_n(\bm{\Psi})\right|^2}{\sigma^2}\right) \quad.
\end{equation}
The joint probability of the full signal vector $\bm{s}$ is:
\begin{equation}
\begin{split}
        p(\bm{s};\bm{\Psi}) & = \prod_{n=1}^{N} p(s_n;\bm{\Psi}) \\ &= \left(\frac{1}{\pi\sigma^2}\right)^N \exp\!\left(-\sum_{n=1}^N\frac{\left|s_n-\mu_n(\bm{\Psi})\right|^2}{\sigma^2}\right) \quad .
\end{split}
\end{equation}

The Fisher Information Matrix  is given by
\begin{equation}\label{eq:FIM}
    F_{ij}(\bm{\Psi}) = -\mathrm{E}\left[\frac{\partial^2\log p}{\partial\Psi_i\partial\Psi_j}\right] \quad,
\end{equation}
where $\log p = \log\big(p(\bm{s};\bm{\Psi})\big)$ is the log-likelihood function:
\begin{equation}
    \log p = -N\log(\pi\sigma^2) - \sum_{n=1}^N \frac{\left|\varepsilon_n(\bm{\Psi})\right|^2}{\sigma^2}\quad,
\end{equation}
with $\varepsilon_n(\bm{\Psi}) \equiv s_n - \mu_n(\bm{\Psi})$ the residual. Using $\partial s_n/\partial\Psi_i = 0$, since $s_n$ is fixed observed data, we obtain
\begin{equation}\label{eq:depsdpsi}
    \frac{\partial \varepsilon_n(\bm{\Psi})}{\partial\Psi_i} = -\frac{\partial\mu_n(\bm{\Psi})}{\partial\Psi_i} \quad,
\end{equation}
and hence
\begin{equation}\label{eq:dabs2}
    \frac{\partial \left|\varepsilon_n(\bm{\Psi})\right|^2}{\partial\Psi_i} = 2\,\mathrm{Re}\!\left(\varepsilon_n^*(\bm{\Psi})\,\frac{\partial\mu_n(\bm{\Psi})}{\partial\Psi_i}\right)\quad .
\end{equation}

The score function, \textit{i.e.} the derivative of the log-likelihood, can then be written as
\begin{equation}
    \frac{\partial\log p}{\partial\Psi_i} = \frac{2}{\sigma^2} \sum_{n=1}^N\mathrm{Re}\!\left(\varepsilon_n^*(\bm{\Psi})\,\frac{\partial \mu_n(\bm{\Psi})}{\partial\Psi_i}\right) \quad .
\end{equation}

Differentiating once more, with respect to $\Psi_j$:
\begin{equation}\label{eq:hessian}
\begin{split}
    &\frac{\partial^2\log p}{\partial\Psi_i\partial\Psi_j}
    = \frac{2}{\sigma^2} \sum_{n=1}^N\frac{\partial}{\partial\Psi_j}\,\mathrm{Re}\!\left(\varepsilon_n^*(\bm{\Psi})\,\frac{\partial \mu_n(\bm{\Psi})}{\partial\Psi_i}\right) \\
    &= \frac{2}{\sigma^2} \sum_{n=1}^N\mathrm{Re}\!\left(
-\frac{\partial\mu_n^*(\bm{\Psi})}{\partial\Psi_j}\,\frac{\partial \mu_n(\bm{\Psi})}{\partial\Psi_i} + \varepsilon_n^*(\bm{\Psi})\,\frac{\partial^2\mu_n(\bm{\Psi})}{\partial\Psi_i\partial\Psi_j}\right) \quad .
\end{split}
\end{equation}

Injecting Eq.~\eqref{eq:hessian} into Eq.~\eqref{eq:FIM}: the first term contains no $\varepsilon_n$ and is built purely from  deterministic $\mu_n(\bm\Psi)$ and its derivatives, hence it passes through $\mathrm{E}[\cdot]$ unchanged. The second term vanishes in expectation because $\mathrm{E}[\varepsilon_n^*(\bm{\Psi})] = 0$ at the true parameter $\bm{\Psi}$ (since $\varepsilon_n(\bm{\Psi}_{\text{true}}) = b_n$, which is zero-mean noise). The Fisher information matrix can therefore finally be written as
\begin{equation}\label{eq:fisher_scalar}
    F_{ij}(\bm{\Psi}) = \frac{2}{\sigma^2}\sum_{n=1}^N \mathrm{Re}\left(\frac{\partial \mu_n(\bm{\Psi})}{\partial\Psi_i}\left(\frac{\partial \mu_n(\bm{\Psi})}{\partial\Psi_j}\right)^*\right) \quad .
\end{equation}

In matrix form:
\begin{equation}\label{eq:fisher_matrix}
    \bm{F} = \frac{2}{\sigma^2}\,\mathrm{Re}\!\big(\bm{G}^H\bm{G}\big), \qquad
    \bm{G} = \left[\frac{\partial\bm{\mu}}{\partial \Psi_1},\, \ldots,\, \frac{\partial\bm{\mu}}{\partial \Psi_M}\right] \quad,
\end{equation}
where $M$ is the number of parameters.

\subsection{Single lossless wave}

Considering a single wavenumber with no attenuation ($k_1''=0$), the only parameter is the real part $k_1'$. Differentiating $\mu_n = a_1 e^{ik_1' x_n}$,
\begin{equation}
    \frac{\partial \mu_n}{\partial k_1'} = a_1\,(i x_n)\,e^{ik_1' x_n}
    = a_1\,[\bm{D}_1]_n,\qquad \bm{D}_1 = i\bm{X}\bm{v}_1,\quad \bm{X}=\mathrm{diag}(x_n),
\end{equation}
so $\bm{G}=a_1\bm{D}_1\in\mathbb{C}^{N\times 1}$. The Fisher information is the
scalar
\begin{equation}
    F = \frac{2}{\sigma^2}\,\Re\big(\bm{G}^H\bm{G}\big)
    = \frac{2|a_1|^2}{\sigma^2}\,\|\bm{D}_1\|^2
    = \frac{2|a_1|^2}{\sigma^2}\sum_{n=1}^N x_n^2,
\end{equation} 
using $\|\bm{D}_1\|^2=\sum_n x_n^2|[\bm{v}_1]_n|^2=\sum_n x_n^2$ since
$|[\bm{v}_1]_n|=1$ in the lossless case. Taking the inverse,
\begin{equation}
    \mathrm{CRB}(k_1') = F^{-1}
    = \frac{\sigma^2}{2|a_1|^2\sum_{n=1}^N x_n^2}.
\end{equation}
For a uniform array $x_n=n\Delta x$, $n=1,\ldots,N$, $\sum_{n=1}^N x_n^2\approx \Delta x^2 N^3/3$ for large $N$, giving
\begin{equation}
    \mathrm{CRB}(k_1')\approx \frac{3\sigma^2}{2|a_1|^2 N^3\Delta x^2}.
\end{equation}
Written as a relative error, with aperture $L\approx N\Delta x$ and
$\mathrm{SNR}=|a_1|/\sigma$,
\begin{equation}
   \frac{\sqrt{\mathrm{CRB}(k_1')}}{k_1'}
    \approx \sqrt{\frac{3}{2}}\frac{1}{\mathrm{SNR}\cdot k_1' L\cdot \sqrt{N}}.
\end{equation}

\subsection{Single attenuated wave}

Now reinstating the imaginary part, the same wave has two parameters,
$k_1'$ and $k_1''$, and its propagation term now carries a decay,
$[\bm{v}_1]_n=e^{ik_1' x_n}e^{-k_1'' x_n}$. Differentiating $\mu_n = a_1e^{ik_1' x_n}e^{-k_1'' x_n}$ with respect to $k_1'$ and $k_1''$:
\begin{equation}
    \frac{\partial \bm{\mu}}{\partial k_1'} = a_1\bm{D}_1,\qquad
    \frac{\partial \bm{\mu}}{\partial k_1''} = i\,a_1\bm{D}_1,\qquad
    \end{equation}
so $\bm{G}=[\,a_1\bm{D}_1,\ i a_1\bm{D}_1\,] \in\mathbb{C}^{N\times 2}$ which leads to:
\begin{equation}
    \bm{G}^H\bm{G} = |a_1|^2\|\bm{D}_1\|^2
    \begin{bmatrix} 1 & i\\ -i & 1\end{bmatrix}, \quad \|\bm{D}_1\|^2=\sum_{n=1}^N x_n^2\,e^{-2k_1'' x_n} \quad .
\end{equation}

On taking $\tfrac{2}{\sigma^2}\Re(\cdot)$ the off-diagonal entries
($\Re\{\pm i\}=0$) vanish, leaving a diagonal Fisher matrix proportional to the identity:
\begin{equation}
    \bm{F} = \frac{2|a_1|^2}{\sigma^2}\,\sum_{n=1}^N x_n^2\,e^{-2k_1'' x_n}\begin{bmatrix} 1 & 0\\ 0 & 1\end{bmatrix}
    .
\end{equation}
The CRB for each parameter is the inverse of the corresponding diagonal entries of the Fisher matrix, \textit{i.e.} $\text{CRB}(\Psi_i) = [F_{ii}]^{-1}$. Therefore, it gives the same Cram\'er-Rao bound for the real and imaginary part of the wavenumber:
\begin{equation}
    \mathrm{CRB}(k_1') = \mathrm{CRB}(k_1'')
    = \frac{\sigma^2}{2|a_1|^2\sum_{n=1}^N x_n^2\,e^{-2k_1'' x_n}}.
\end{equation}

\bibliography{bibref}

\end{document}